\newcommand{\un}{~\mathrm} 
\newcommand{\ie}{{\em i.e. }} 
\newcommand{\eg}{{\em e.g. }}

\RequirePackage{fix-cm}
\documentclass[twocolumn,final]{svjour3}          
\smartqed  
\usepackage[T1]{fontenc}
\usepackage{graphicx}
\usepackage{natbib} 
\usepackage{amsmath} 
\usepackage{amsfonts}
\usepackage{amssymb}  
\usepackage{amssymb} 
\begin{document}

\title{Damage mechanisms in the dynamic fracture of nominally brittle polymers}

\author{Davy Dalmas \and Claudia Guerra \and Julien Scheibert \and Daniel Bonamy}

\institute{Davy Dalmas \at
              Unité Mixte CNRS/Saint-Gobain, Surface du Verre et Interfaces, 39 Quai Lucien Lefranc, 93303 Aubervilliers cedex, France
           \and Claudia Guerra \at
              CEA, IRAMIS, SPCSI, Grp. Complex Systems $\&$ Fracture, F-91191 Gif sur Yvette, France
           \and Julien Scheibert \at
              CEA, IRAMIS, SPCSI, Grp. Complex Systems $\&$ Fracture, F-91191 Gif sur Yvette, France \\
              Laboratoire de Tribologie et Dynamique des Systèmes, CNRS, Ecole Centrale de Lyon, 36 Avenue Guy de Collongue, 69134 Ecully Cedex, France
              \and Daniel Bonamy \at
              CEA, IRAMIS, SPCSI, Grp. Complex Systems $\&$ Fracture, F-91191 Gif sur Yvette, France \\
              \email{Daniel.Bonamy@cea.fr}           
}

\date{Received: date / Accepted: date}

\maketitle

\begin{abstract}
Linear Elastic Fracture Mechanics (LEFM) provides a consistent framework to evaluate quantitatively the energy flux released to the tip of a growing crack. Still, the way in which the crack  selects  its  velocity  in  response  to  this  energy  flux  remains  far  from  completely  understood. To uncover the underlying mechanisms, we experimentally studied damage and dissipation processes that develop during the dynamic failure of polymethylmethacrylate (PMMA), classically considered as the archetype of brittle amorphous materials. We evidenced a well-defined critical velocity along which failure switches from nominally-brittle to quasi-brittle, where crack propagation goes hand in hand with the nucleation and growth of microcracks.  Via post-mortem analysis of the fracture surfaces, we were able to reconstruct the complete spatiotemporal microcracking dynamics with micrometer/nanosecond resolution. We demonstrated that the true local propagation speed of individual crack fronts is limited to a fairly low value, which can be much smaller than the apparent speed measured at the continuum-level scale. By coalescing with the main front, microcracks boost the macroscale velocity through an acceleration factor of geometrical origin. We discuss the key role of damage-related internal variables in the selection of macroscale fracture dynamics.
\end{abstract}

\section{Introduction}\label{intro}

Dynamic crack propagation drives catastrophic material failure and is usually described using the Linear Elastic Fracture Mechanics (LEFM) framework \citep{Freund90_book,Ravichandar04_book}. This theory considers the straight propagation of a single smooth crack and assumes all energy dissipating processes to be entirely localized in a small zone at the crack tip, so called fracture process zone (FPZ). Crack growth velocity $v$ is then selected by the balance between the mechanical energy that flows within the FPZ per time unit and the dissipated energy within the FPZ over the same time unit. This yields \citep{Freund90_book}:

\begin{equation}
\Gamma \simeq \left( 1-v/C_R \right) {K^2(c)}/{E}, 
\label{Eq1}
\end{equation}

\noindent where $C_R$ and $E$ are the Rayleigh wave speed and the Young modulus of the material, respectively, $\Gamma$ is the fracture energy, and $K(c)$ is the Stress Intensity Factor (SIF) for a quasi-static crack of length $c$. $K$ only depends on the applied loading and the cracked specimen geometry, and entirely characterizes the stress field in the vicinity of the crack front.
 
LEFM predictions agree well with observations as long as the crack growth is sufficiently slow \citep{Bergkvist74_efm}. However, large discrepancies are reported at high speed \citep{Ravichandar04_book,Bouchbinder10_arcms}. In particular, the maximal crack speeds attained experimentally in amorphous materials are far smaller (typically by a factor of two) than the limiting speed $C_R$ predicted by Eq.~\ref{Eq1}.
 
The existence of a micro-branching instability \citep{Fineberg92_prb} at a critical velocity $v_{b}$ (typically of the order of $0.35 - 0.4 C_R$) permits to explain {\em part} of this discrepancy: Beyond $v_{b}$, the crack front splits into a multiple crack state that cannot be described by Eq.~\ref{Eq1} anymore. However, it is commonly stated \citep{Sharon99_nature} that LEFM works well below $v_b$, or more generally when microbranching instabilities are absent. As a matter of fact, most recent works focused on failure regimes beyond $v_b$ (see e.g. \citep{Gumbsch97_prb,Addabedia99_prl,Henry04_prl,Bouchbinder05_pre,Spatschek06_prl,Henry08_epl}). Still, several observations reported at lower speeds remain puzzling. In particular, even for velocities much lower than $v_b$, the measured dynamic fracture energy is generally found to be much higher than that at crack initiation \citep{Sharon99_nature,Kalthoff76_ijf,Rosakis84_jmps,Fond01_cras}.
 
The experiments reported here were designed to better understand the mechanisms that select crack velocity in dynamic fracture. In this context, we developed an experimental setup (presented in section 2) that permits to characterize over a wide range of crack speeds, in particular speeds below $v_{b}$, the dissipative and damage processes that develop in polymethylmethacrylate (PMMA), classically considered as the archetype of brittle amorphous materials. This setup allowed us to evidence a novel critical velocity, smaller than $v_b$, beyond which crack propagation goes hand in hand with the nucleation and growth of microcracks ahead of the main crack front (section 3).  Via accurate {\em post-mortem} analysis of the patterns let on fracture surfaces, we were able to reconstruct the full spatio-temporal dynamics of these microfailure events (section 4). Their statistics and their dependency with crack tip loading have been characterized (section 5). In section 6, we will show how microcracking acts and selects the apparent velocity measured at the continuum-level scale.

\section{Experimental setup}\label{sec:2}

\begin{table*}
\begin{center}
\begin{tabular} {l c c c c c c}
	\hline
	& $E$(GPa)& $\nu $ & Density (kg/dm$^3$)& $C_d$ (m/s)& $C_s$ (m/s)& $C_R$ (m/s)\\ 
    \hline
    PMMA & $2.8 \pm 0.2$ & $0.36 \pm 0.08$ & $1.18 \pm 0.02$ & $2010 \pm 60$ & $950 \pm 30$ & $880 \pm 20$\\
    \hline
\end{tabular}
\end{center}
\caption {Mechanical properties of the PMMA used in our fracture experiments. $E$ and$\nu$ denote the Young modulus and the Poisson ratio, respectively. $C_D$, $C_S$, and $C_R$ denote the speeds of dilational, shear and Rayleigh waves respectively.}
\label{table2-1}
\end{table*}

\begin{figure}
\centering
\includegraphics[width=\columnwidth]{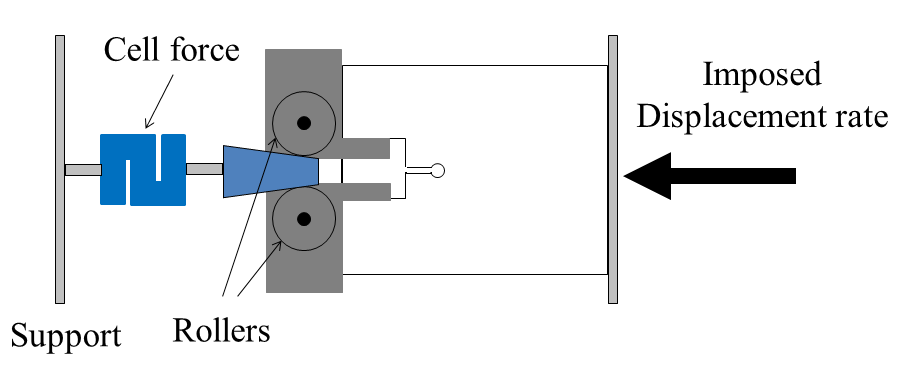}
\includegraphics[width=0.9\columnwidth]{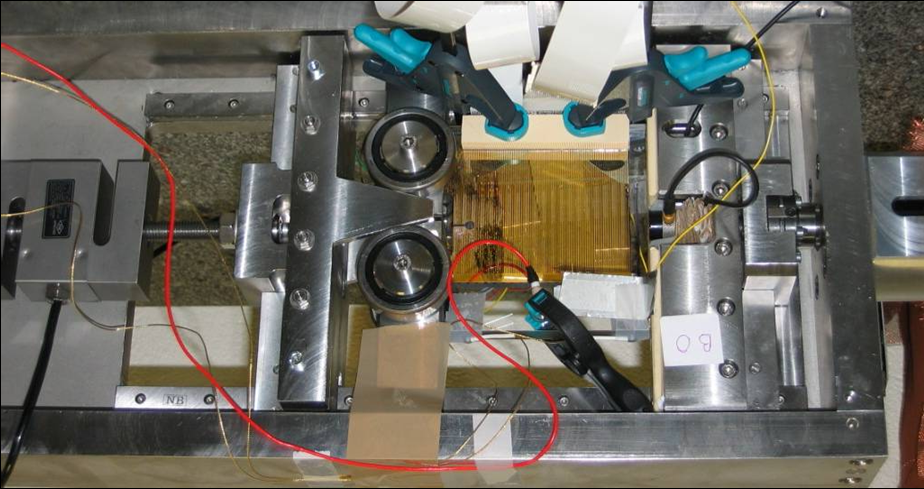}
\caption{Sketch (top) and photo (bottom) of the experimental setup.}
\label{Fig2-1}
\end{figure}

\begin{figure*}
\centering
\includegraphics[width=1.7\columnwidth]{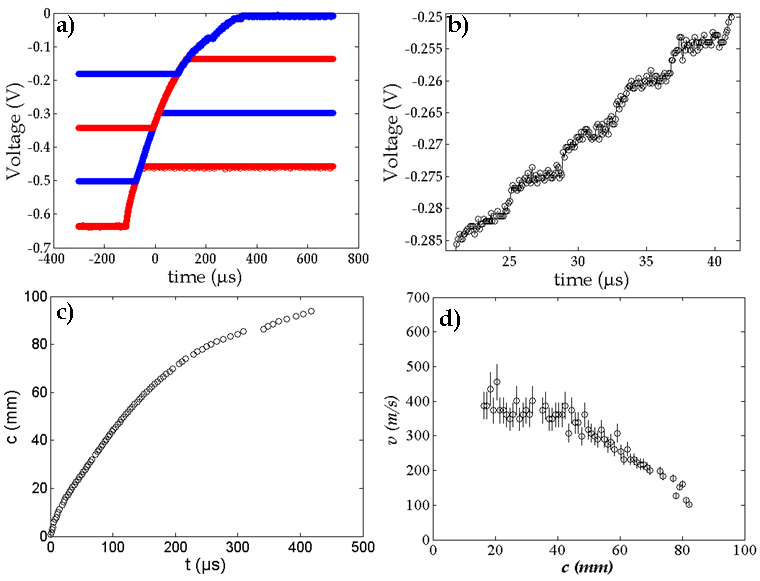}
\caption{Measurement of instantaneous crack velocity. a: Voltage as a function of time as recorded via the oscilloscope. b: Zoom that permits to distinguish the individual jumps yielded by the successive cuts of the conductive lines as crack advances. c: Resulting position of crack tip as a function of time. d: Final curve showing the variations of crack velocity with crack length.}
\label{Fig2-2}
\end{figure*}

\begin{figure*}
\centering
\includegraphics[width=1.3\columnwidth]{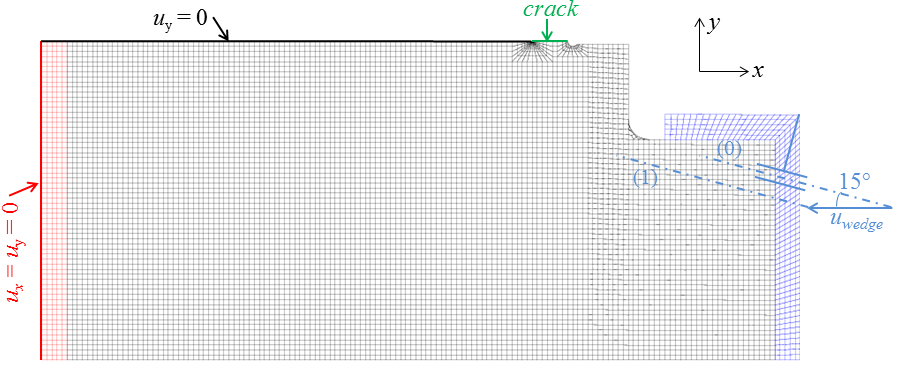}\\
\includegraphics[height=0.35\columnwidth]{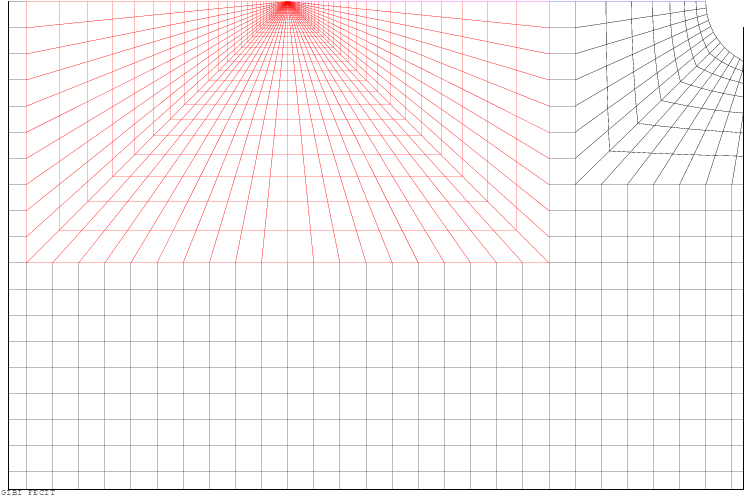}
\includegraphics[height=0.35\columnwidth]{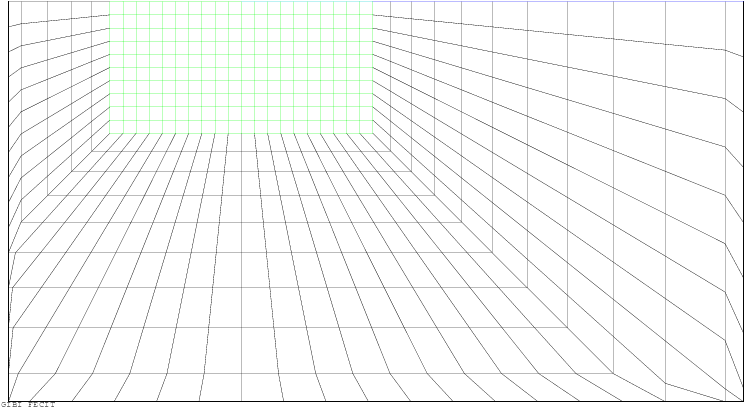}
\caption{Computation of SIF. Top: Typical mesh used for finite elements calculations, in order to access the stress/strain fields in the experiments. Red: polymeric layer. Black: sample. Blue: L-shaped block. Green line: cracked line. The slide link connected to the L-shaped block is used to model the motion of the contact point between the pushed wedge and the roller. Specimen loading is achieved by translating horizontally the slide link, from (0) to (1), over a distance $u_{wedge}$ selected so that the horizontal force applying on the slide link is half that measured experimentally. Bottom-Left: Zoom on the meshing in the transition region (red) between coarse meshing ($1~\un{mm}$ mesh size) in the bulk and fine meshing close to the crack tip. On the right is shown part of the circular hole at the seed crack tip. Bottom-Right: Zoom on the crack-tip region (green), meshed with a size of $1\un{nm}$. }
\label{Fig2-3}
\end{figure*}

Fracture tests were performed in the wedge splitting geometry \citep{Bruhwiler90_efm,Karihaloo01_ijf} depicted in Fig. 1. Specimens were prepared from PMMA parallelepipeds of size $140 \times 125 \times 15\un{ mm}^3$ in the propagation ($x$-axis), loading ($y$-axis), and thickness ($z$-axis) directions, respectively. Table 1 gives the main characteristics of the PMMA used in this study.  Subsequently, a notch was shaped (i) by cutting a $25\times25\un{mm}^2$ rectangle from the middle of one of the $125 \times 15\un{mm}^2$ edges; and (ii) by subsequently adding a 10-mm groove deeper into the specimen. Two steel blocks equipped with rollers were then placed on both sides of this notch and the specimen was loaded by pushing a wedge (semi-angle of $15^\circ$) at a small constant velocity ($40~\mu\mathrm{m/s}$) between these two blocks. This permits (i) to spread the loading force over a large contact area and prevent any plastic deformation of PMMA at the loading contacts; and (ii) to suppress friction in the system. As a result, the vicinity of the crack tip can be assumed to be the sole dissipation source for mechanical energy in the system.

In such a wedge splitting geometry, the SIF decreases with crack length. To obtain dynamic failure, we then introduced a circular hole of tunable diameter from $2\un{mm}$ to $8\un{mm}$ at the tip of the seed crack. This hole blunts the seed crack and delays the propagation onset. This increases the mechanical energy stored in the sample at the onset of crack propagation. It allowed us to widen the range of SIF (from $0.9~$MPa.$\sqrt{\mathrm{m}}$ to $4.5$MPa.$\sqrt{\mathrm{m}}$) and that of crack velocity (from $75\un{m/s}\approx 0.08C_R$ to $500\un{m/s}\approx 0.57C_R$) accessible in the experiments. 

The apparatus used to carry out the experiments is homemade (Fig. \ref{Fig2-1}). One side of the machine is fixed while the other moves via a stepper motor (Oriental motor EMP400 Series) allowing incremental displacements as small as $40\un{nm}$. The compressive force is measured, up to $20\un{kN}$ via a S-type Vishay load cell 363 Series. In all performed tests, both the displacement and the force are recorded via a computer with a time resolution of $1\un{s}$. In particular, the value of the applied force at propagation onset is recorded and later used in the finite element analyses described thereafter.

To measure the instantaneous crack velocity, we used a modified version of the electrical resistance grid technique \citep{Cotterell68_ijf,Anthony70_pm,Boudet96_jpii,Fineberg92_prb}. A series of 90 parallel conductive lines (2.4-nm-thick Cr layer covered with 23-nm-thick Au layer) were deposited on one of the two $140 \times 125\un{mm}^2$ sides of the specimen using magnetron sputtering deposition through a polymer mask glued on the sample surface. Stripes width and length are $500\mu\mathrm{m}$ and $100\un{mm}$, respectively, and they are separated by gaps of $500\mu\mathrm{m}$. Once the sample is installed in the compression machine, each line is connected in series with a resistor of $ 30\un{k}\Omega$ -- this value was chosen to be extremely high with respect to that of the strip (about few ohms) so that the latter is negligible. As the crack propagates throughout the specimen, it successively cuts the conductive lines, which, each time, makes the global system resistance increase by a constant increment. The time locations of these jumps are detected via an oscilloscope (acquisition rate: $10\un{MHz}$) through a voltage divider circuitry. In order to increase the accuracy of jump detection, we used the four channels of the oscilloscope in cascade with different offsets, so that each channel is triggered as the previous one gets out of its range (Fig. \ref{Fig2-2}:A and B). As a result, we obtained the crack length $c(t)$ as a function of time with space and time accuracies of $40~\mu\mathrm{m}$ and $0.1\mu\mathrm{s}$, respectively (Fig. \ref{Fig2-2}:C). The profile of instantaneous velocity $v(c)$ as a function of crack length can then be easily deduced (Fig. \ref{Fig2-2}:D).

Finite element analysis was used to estimate the SIF evolution during failure. Figure \ref{Fig2-3}:Top represents the meshing of the complete system. The average mesh size is $1\un{mm}$ and reduces logarithmically down to $1\un{nm}$ at the crack tip (Fig. \ref{Fig2-3}:Bottom), in order to resolve in space the crack tip opening. Only half of the system is required for symmetry reasons. Young modulus and Poisson ratio are set to $2.8\un{GPa}$ and $0.36$, respectively (see Tab. 1). The boundary conditions are the following: (i) The left edge is perfectly adhesive to a polymeric layer of thickness $5\un{mm}$, Young's modulus $3\un{GPa}$ and Poisson ratio $0.41$; (ii) The left edge of the layer has a no-displacement condition; (iii) On the right, the notch is loaded through a L-shaped block, of thickness $5\un{mm}$, Young modulus $300\un{GPa}$ and Poisson ratio 0.4, perfectly adhering to the sample; (iv) The L-shaped block is bounded to a virtual roller, the center of which can only move along a line parallel to the wedge side, at a distance equal to the roller radius; (v) The crack edge is stress free; and (vi) Vertical displacement is forbidden along the uncracked part of the symmetry plane (top line of Fig. \ref{Fig2-3}:Top). For each sample, the equilibrium position of the wedge yielding the measured applied load at the onset of crack propagation was determined using a plane stress static finite element code (Cast3M). Quasi-static crack propagation is then simulated by increasing the crack length while imposing the wedge position to remain constant. This latter assumption is experimentally justified. Indeed, it has been observed that the load cell signal started to be modified only after a few hundreds of microseconds (typical time of an experiment) after crack initiation\footnote{To measure this time shift, we directly connected the Wheatstone bridge of the load cell to an oscilloscope, without passing through the signal conditioner. This latter, indeed, imposes a time resolution of $1\un{s}$.}. This is attributed to the time required for the sound waves to travel from the crack tip to the load cell. As a result, the wedge position can be assumed to be practically constant during crack propagation. Static SIF is then determined, for each value $c$ of crack length, using the J-integral method \citep{Rice68_jam}. The validity of our SIF calculation was benchmarked against literature results for wedge-splitting configurations: Our code agreed to better than $2\%$ with the results of \citet{Karihaloo01_ijf} when using their system parameters.
 
To image the post-mortem fracture surfaces, we used a Leica DM2500 microscope. Most of the images were taken with a $\times 5$ objective under polarized light. The resulting images are $1.4\times 1\un{mm}^2$ in area, with a pixel size of $677\un{nm}$. We also imaged the fracture surfaces with an interferometry profilometer (FOGALE Nanotech) which allows one to gather topographical information. These 3D images were taken with a $ \times 5$ objective. Their area is $1.4\times 1\un{mm}^2$ with a pixel size of $1.86~\mu\mathrm{m}$.

\section{Selection of fracture energy with velocity and evidence of a brittle/quasibrittle transition}\label{sec:3}

\begin{figure*}
\centering
\includegraphics[width=0.64\columnwidth]{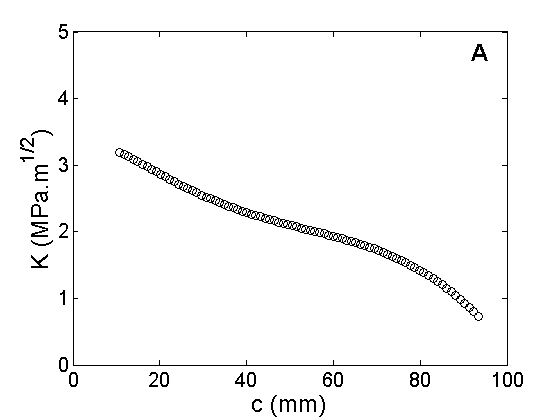}
\includegraphics[width=0.64\columnwidth]{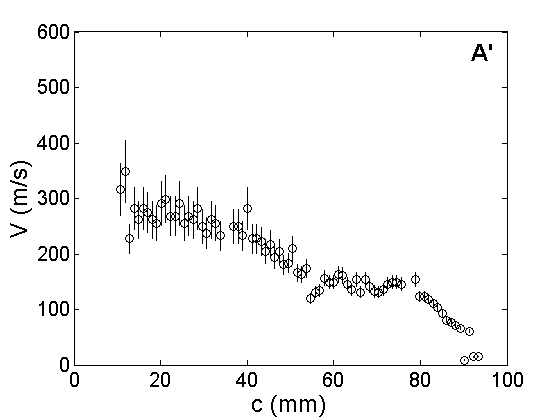}
\includegraphics[width=0.64\columnwidth]{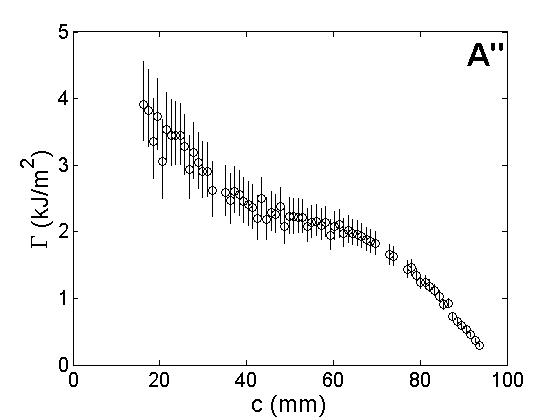}
\includegraphics[width=0.64\columnwidth]{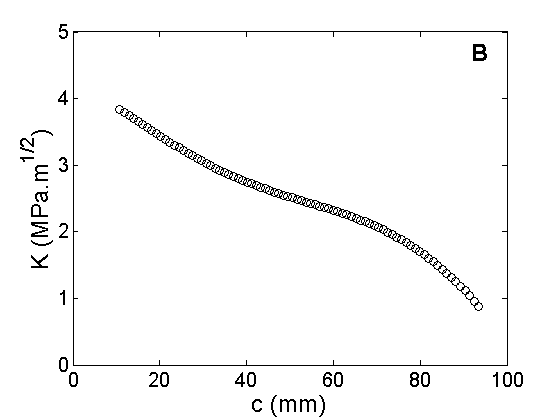}
\includegraphics[width=0.64\columnwidth]{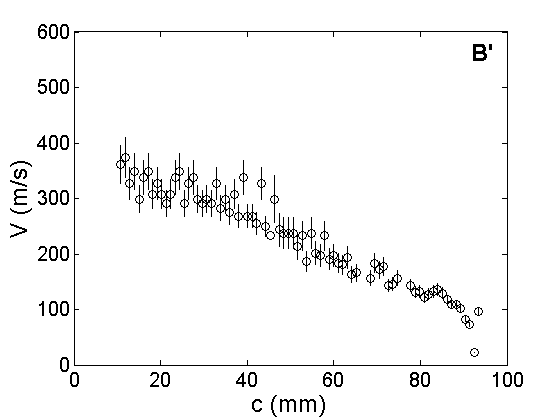}
\includegraphics[width=0.64\columnwidth]{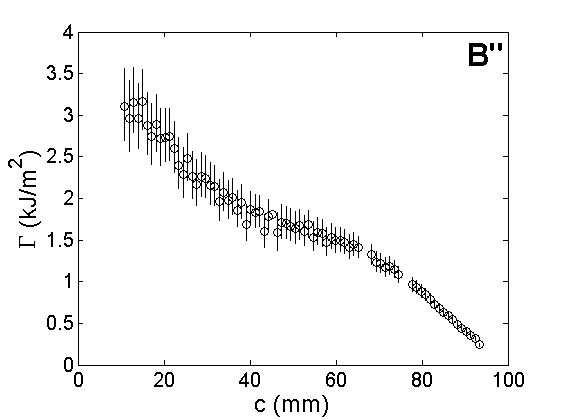}
\includegraphics[width=0.64\columnwidth]{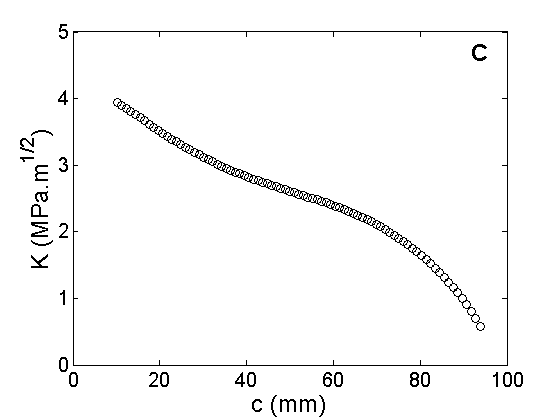}
\includegraphics[width=0.64\columnwidth]{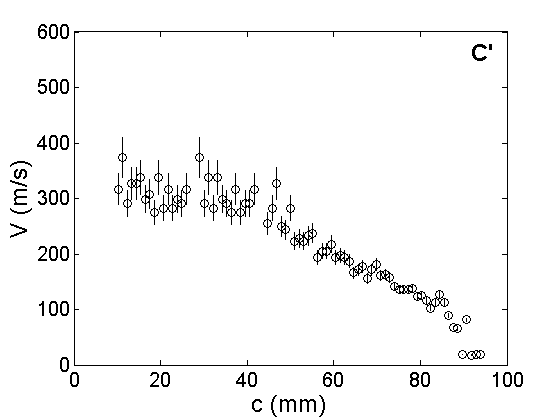}
\includegraphics[width=0.64\columnwidth]{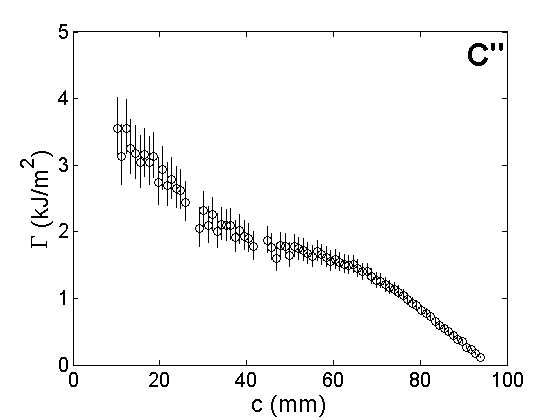}
\includegraphics[width=0.64\columnwidth]{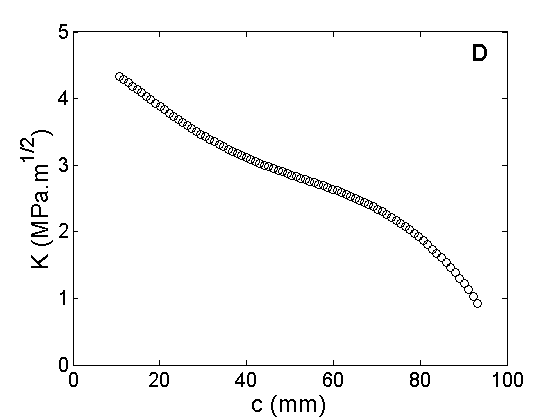}
\includegraphics[width=0.64\columnwidth]{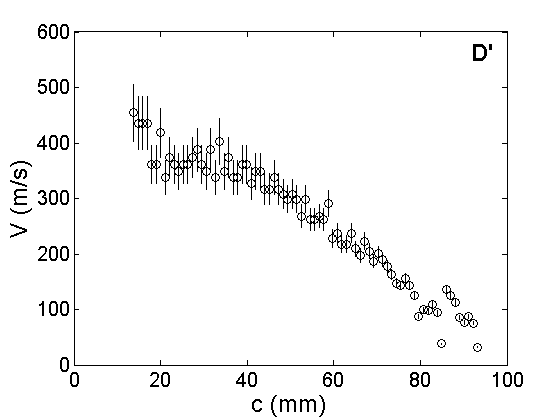}
\includegraphics[width=0.64\columnwidth]{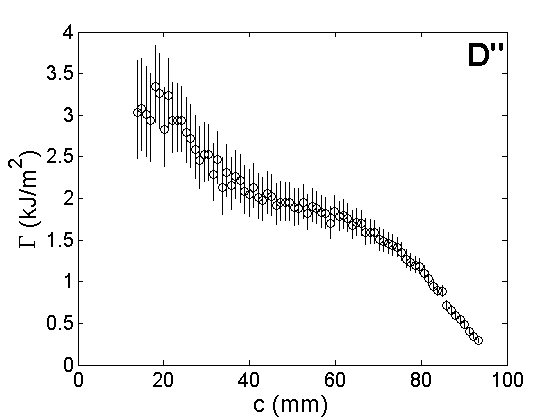}
\includegraphics[width=0.64\columnwidth]{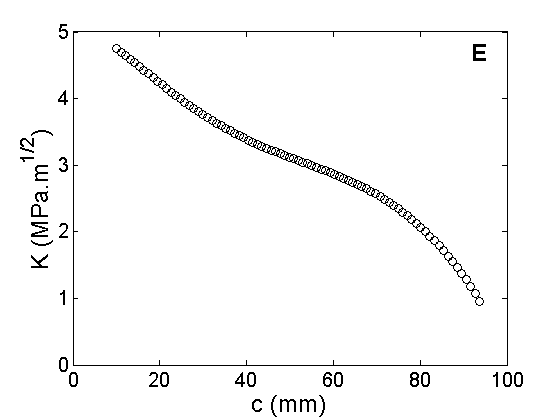}
\includegraphics[width=0.64\columnwidth]{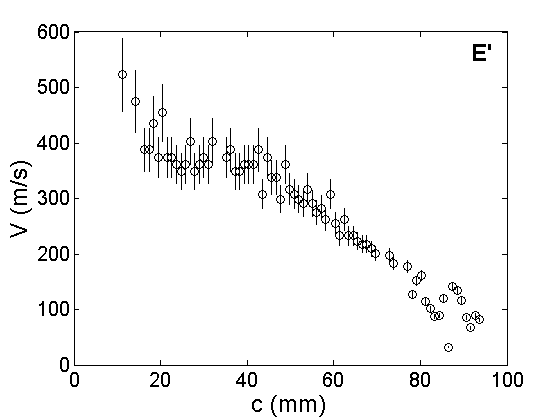}
\includegraphics[width=0.64\columnwidth]{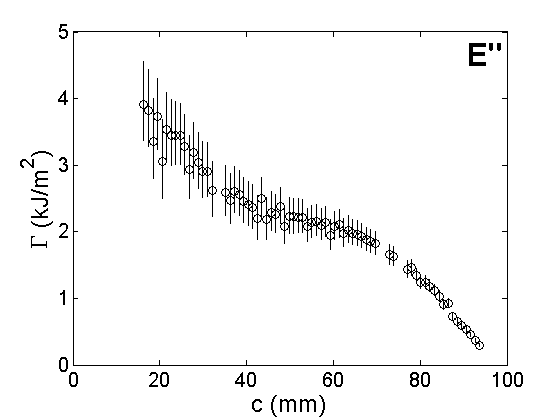}
\caption{Variation of SIF $K$ (left), instantaneous crack velocity $v$ (center) and fracture energy $\Gamma$ (right) as a function of crack length $c$ for five experiments with different stored mechanical energy $U_0$ at propagation onset: $2.0\un{J}$ (A, A' and A''), $2.6\un{J}$ (B, B' and B''), $2.9\un{J}$ (C, C' and C''), $3.8\un{J}$ (D, D' and D''), and $4.2\un{J}$ (E, E' and E'').}
\label{Fig3-1}
\end{figure*}

\begin{figure}
\centering
\includegraphics[width=0.9\columnwidth]{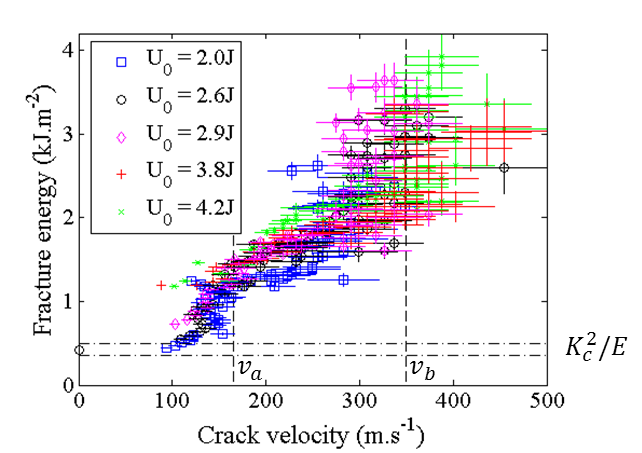}
\caption{Fracture energy $\Gamma$ as a function of crack velocity $v$ for five different experiments with different stored mechanical energies $U_0$ at crack initiation. The two vertical dashed lines correspond to $v_a$ and $v_b$. The two horizontal dashed lines indicate the confidence interval ($95\%$) for the measured fracture energy $K_c^2/E$ at crack initiation (Taken from \citet{Scheibert10_prl}).}
\label{Fig3-2}
\end{figure}

In each of the experiments performed, both the SIF profile $K(c)$ and the instantaneous velocity profile $v(c)$ have been determined (Figs. \ref{Fig3-1}:Left and center). From these curves, one can derive the profiles of fracture energy, $\Gamma(c)$, using Eq. \ref{Eq1} (Fig. \ref{Fig3-1}:Right). We note here that Eq. \ref{Eq1} is strictly valid in the case of crack propagation in an elastic half-space, i.e. in the absence of waves emitted by the crack and coming back to it after reflection on sample boundaries (\cite{Freund90_book,Goldman10_prl}). We argue that our experiments are essentially unaffected by such reflected wave for at least two reasons: i) as measured by \citet{Boudet95_epl}, the sound energy emitted by a crack in PMMA lies within the range $1 - 4\un{J/m}^2$, a very small value compared to the energy required to fracture the material (a few $\un{kJ/m}^2$, see below); and ii) this radiated energy is quickly dissipated within the material's bulk -- The acoustic attenuation coefficient has been measured to be $0.67\un{dB/(cm.MHz)}$.

Figure \ref{Fig3-2} superimposes the various curves $\Gamma$ vs. $v$ measured in the various experiments performed at various levels $U_0$ of the initially stored mechanical energy. Several regimes can be evidenced:
\begin{itemize}
\item For small $v$, $\Gamma$ roughly remains constant, close to $K_c^2/E$, as expected within standard LEFM;
\item As v increases and reaches a first critical velocity $v_a \approx 165\un{m/s} \approx 0.19 C_R$, $\Gamma$ suddenly increases to a value about four times larger than $\Gamma(v=0)=K_c^2/E$;
\item  Beyond $v_a$, $\Gamma$  slowly increases with $v$;
\item  As $v$ reaches a second critical velocity $v_b \approx 350\un{m/s} \approx 0.36 C_R$, $\Gamma$ starts to increase rapidly again. It seems to diverge at $v\approx 450 \un{m/s} \approx 0.5C_R$.
\end{itemize}

Note also the fairly good collapse of the different curves $\Gamma(v)$ below $v_b$, and the large dispersion above. This suggests that Eq. \ref{Eq1} is relevant for $v \leq v_b$, -- provided a suitable velocity dependence $\Gamma(v)$ is ascribed--, but not beyond. The second critical value $v_b \approx 350\un{m/s} \approx 0.36 C_R$ is found to correspond to the onset of microbranching instability widely discussed in the literature (see e.g. \citep{Fineberg92_prb,Sharon99_nature}). The first critical value $v_a \approx 165\un{m/s} \approx 0.19 C_R$ was observed in our series of experiments and reported in \citep{Scheibert10_prl} for the first time. This observation was made possible by the use of the wedge-splitting geometry, which is a decelerating crack configuration. It offers many data points at relatively low velocities, contrary to most other devices used before.  The rapid increase of $\Gamma$ with $v$ around $v_a$ provides a direct interpretation for the repeated observations of cracks that span a large range of $\Gamma$, while keeping a nearly constant velocity about $0.2 C_R$ (see e.g. \citep{Ravichandar84b_ijf,Ravichandar97_jmps}).  

\begin{figure*}
\centering
\includegraphics[width=1.8\columnwidth]{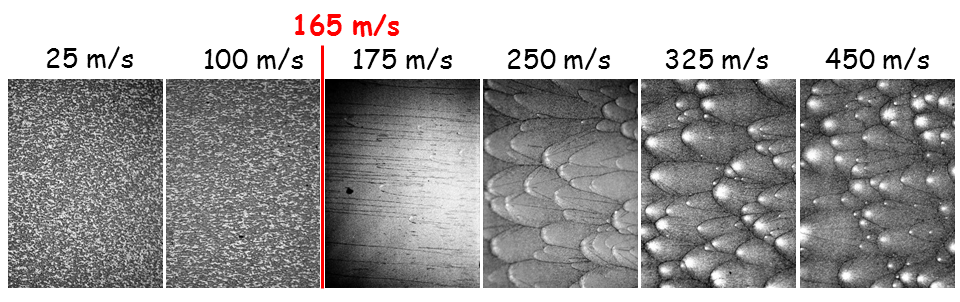}
\caption{$1 \times 1.4$ microscope images of fracture surfaces created at various speed $v$.}
\label{Fig3-3}
\end{figure*}

\begin{figure*}
\centering
\includegraphics[width=1.8\columnwidth]{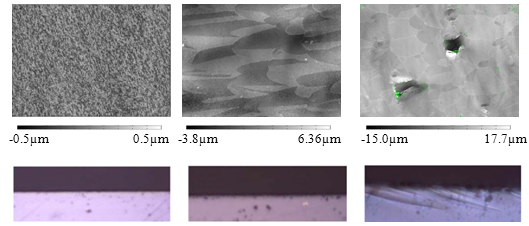}
\caption{Profilometric $1.4\times 1\un{mm}^2$ surface (Top) and optical $0.7\times 0.25 \un{mm}^2$ side-view (Bottom) images taken at velocity smaller than $v_a$ (left), between $v_a$ and $v_b$ (center), and above $v_b$ (right). Crack has propagated from left to right.}
\label{Fig3-4}
\end{figure*}

To shed light on the nature of the transition at $v=v_a$ evidenced on the curve relating $\Gamma$ to $v$, we examined the post-mortem fracture surfaces and their evolution as crack speed increases (Fig. \ref{Fig3-3}). For $v$ smaller than $v_a$, fracture surfaces remain smooth at optical scales. But above this threshold, conic marks start to be observed. These marks remain confined within a roughness scale of few micrometers at the fracture surfaces, contrary to the microbranches that deeply develop into the specimen bulk above $v_b$ (see side views and topographical images in Fig. \ref{Fig3-4}).

Similar conic marks were reported in the fracture of other brittle materials, among which polystyrene \citep{Regel51_ztf}, oxide glasses \citep{Smekal53_oia,Holloway68_pe}, cellulose acetate \citep{Kies50_jap}, polycrystalline materials \citep{Irwin52_wjrs}, or homalite \citep{Ravichandar97_jmps}. They are classically associated \citep{Smekal53_oia,Ravichandar97_jmps,Rabinovitch00_prb} with the presence of microcracks that nucleate and grow ahead of the main crack front, and subsequently coalesce with it. 

\begin{figure}
\centering
\includegraphics[width=0.9\columnwidth]{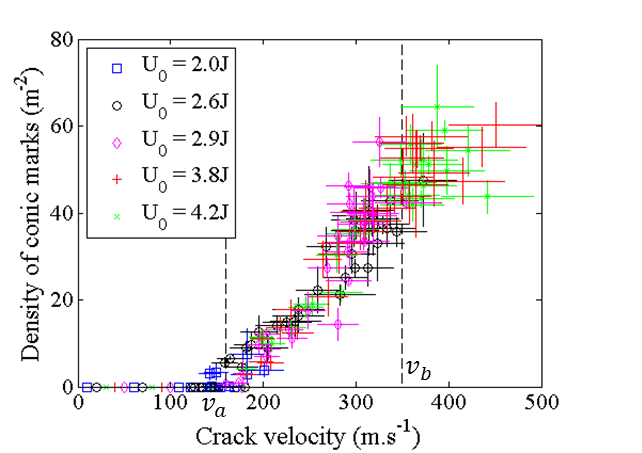}
\caption{Surface density $\rho$ of conic marks as a function of crack velocity for five different experiments with different stored mechanical energies $U_0$ at crack initiation (Taken from \citet{Scheibert10_prl}).}
\label{Fig3-5}
\end{figure}

Figure \ref{Fig3-5} shows the surface density $\rho$ of conic marks as a function of crack speed $v$. Below $v_a$, no mark is observed, irrespectively of the chosen magnification (from $\times 5$ to $\times 50$). Above this value, $\rho$ increases almost linearly with $v - v_a$. The precise correspondence between the critical velocity $v_a$ at which the curve $\Gamma(v)$ exhibits a kink and that at which conic marks start to be observed suggests that both phenomena are the signature of the same transition. Above $v_a$, PMMA failure switches from nominally brittle to quasi-brittle and goes hand in hand with microcracking that develop ahead of the main front. In the following, we will use the term "Damage Zone" (DZ) to refer to the zone where microcracks develop in the vicinity of the main crack tip. We will distinguish this zone from the FPZ (smaller than the DZ) that embeds dissipative mechanisms (crazing for instance) at the tip of each (micro)crack front. 

Note that the region of the fracture surface that bears conic markings also has a characteristic aspect to the naked eye, as it scatters ligth very efficiently. This is markedly different from the high reflectivity of the region that bears no microcrack. In fractography these regions are classically referred to as the mist and mirror zones, respectively (see e.g. \citep{Hull99_book,Rabinovitch08_pre}). Our observation therefore strongly suggests that the mirror/mist transition is simply the morphological counterpart of the brittle/quasi-brittle transition that occurs at $v_a$

\section{Fractographic reconstruction of individual microfailure events}\label{sec:4}

\begin{figure*}
\centering
\includegraphics[width=1.6\columnwidth]{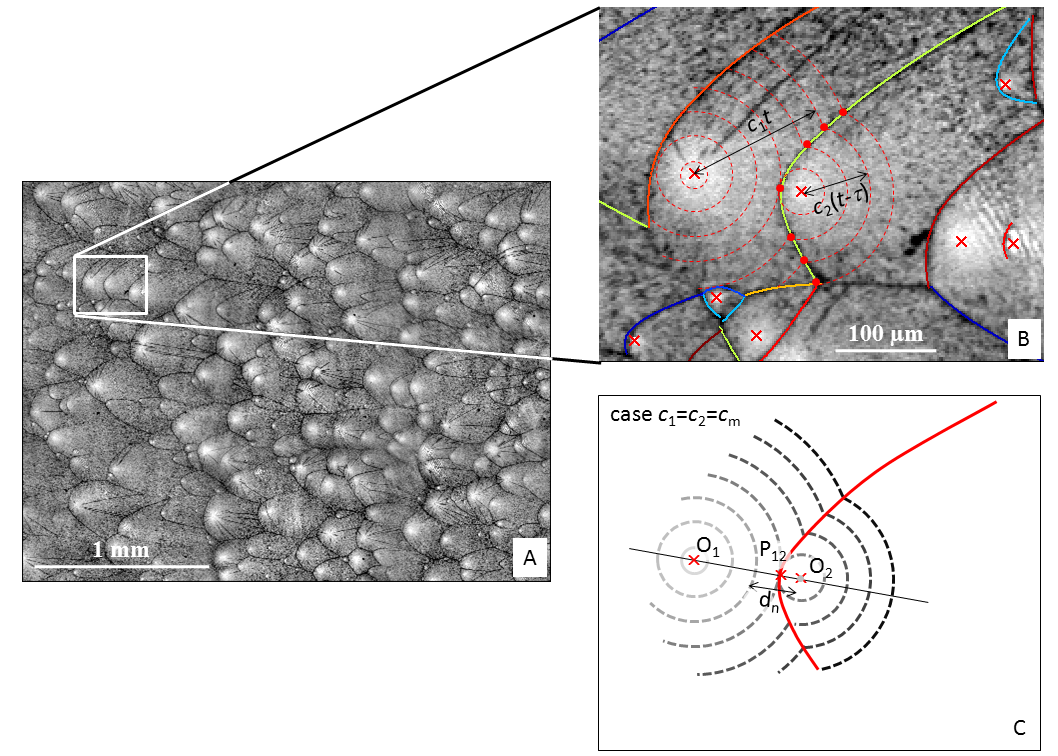}
\caption{A: Typical microscope image of the post-mortem fracture surface. Bright regions indicate the nucleation centers. B: Interpretation of the conic marks: Red dots sketch the successive positions of the fronts that, by interacting, will be giving rise to the green branch of the conic mark. These two microcracks form at $t=0$ and $t=\tau$, then radially grow at speeds $c_1$ and $c_2$. Branch fitting via Eq. \ref{Eq2} permits to measure $c_2/c_1$. Note that a conic mark can be made of several branches. C: When $c_2=c_1$, branches are (mathematical) true conics and the nucleation distance $d_n$ between the front of the incident microcrack (1) and the nucleation point of the forming one (2) is twice the apex-focus distance 0$_2$P$_{12}$ (Taken from \citet{Guerra12_pnas}).}
\label{Fig4-1}
\end{figure*}

In order to further characterize and understand the dynamics of microcracking events which develop during fast fracture in PMMA ($v>v_a$), we proposed a numerical reconstruction based on the post-mortem analysis of the traces left by these events \citep{Guerra12_pnas}. The typical time between two successive microcracking events is about $10\un{ns}$, which makes them inaccessible using standard techniques as e.g. fast imaging or acoustic emission analysis.

The first step is to identify where microcracks have initiated. On the optical images of post-mortem fracture surfaces (Figs. \ref{Fig4-1}:A and B), many well defined points of high optical reflectivity can be seen. These bright/white areas are believed to come from large plastic deformations accompanying the nucleation of microcracks. Thus, they allow us to determine precisely the position of the different nucleation sites. For some microcracks, the precise location of their nucleation site is further constrained by the convergence of fragmentation lines onto it.

The second step is to infer, from the conic patterns, the velocities at which the various microcracks have grown. It is commonly admitted in the literature\citep{Smekal53_oia,Ravichandar97_jmps,Rabinovitch00_prb} that the conical marks indicate the intersection points between two interacting microcracks. Let us then consider two microcracks that propagate radially along slightly different planes with velocities $c_1$ and $c_2$ (Fig. \ref{Fig4-1}:B). The intersection between the initial (micro)crack front and the secondary microcrack leaves a visible trace (i.e. a tiny height difference) on the fracture surfaces. In the coordinate system $(\vec{e}_x,\vec{e}_y)$ centered on the site of nucleation of the first microcrack and chosen so that the $x$-axis passes through the two centers of nucleation, the equation describing the mark left postmortem can be written as \citep{Guerra12_pnas}:

\begin{align}
\frac{y}{\Delta} = \pm &  \left[ \frac{2c^2}{(c^2-1)^2}\frac{c_1\tau}{\Delta}\sqrt{ \left(1-\frac{2x}{\Delta} \right)(c^2-1)+c^2\left(\frac{c_1\tau}{\Delta}\right)^2} \right. \nonumber \\
                       & \quad - \left(\frac{x}{\Delta}\right)^2+\frac{1}{(c^2-1)^2} \nonumber \\
                       & \left. \quad +\frac{c^2+1}{(c^2-1)^2}\left( c^2\left( \frac{c_1\tau}{\Delta}\right)^2 -\frac{2 x}{\Delta}\right) \right]^{1/2},
\label{Eq2}
\end{align}

\noindent where $c = c_2/c_1$, $\Delta$ is the distance between the two nucleation sites, and $\tau$ is the time interval between two microcracking events. The analysis of this equation shows that the number $c_1 \tau / \Delta$ sets the aspect ratio of the fractographic trace, while the ratio $c = c_2/c_1$ fixes its form (see \cite{Guerra12_pnas} for details).

\begin{figure*}
\centering
\includegraphics[width=1.8\columnwidth]{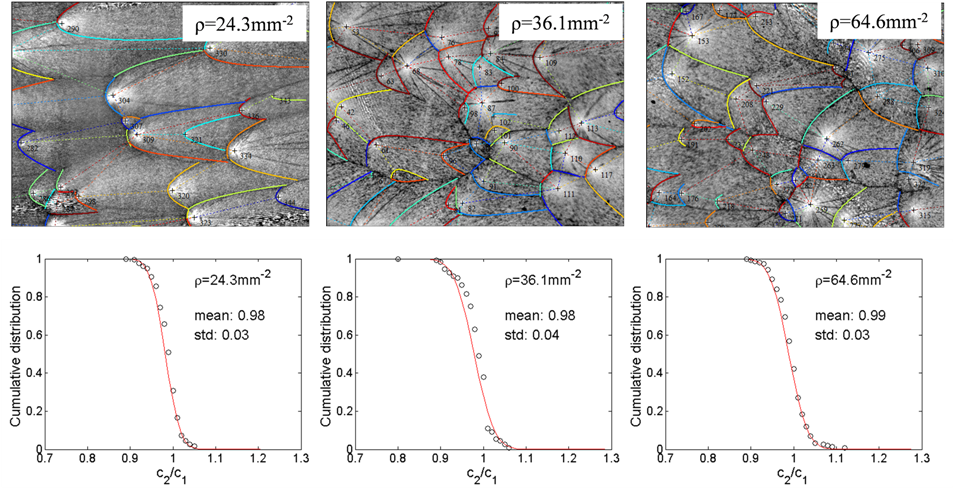}
\caption{Direct extraction of the relative speed between two interacting microcracks at three different microcrack densities. (Top) Zones of investigation. Each conic branch has been attributed a given color and the nuclei of the two corresponding interacting microcracks has been joined by a dotted segment of the same color. Note that a conic-like mark is often made of several of these conic branches. The ratio $c_2/c_1$ is the only adjustable parameter in equation Eq. \ref{Eq2} to determine the branch geometry once the nuclei position and the branch apex are set. (Bottom) Corresponding distributions for $c_2/c_1$: In the three cases, the distributions are found to fit normal distributions of mean value $0.98 - 0.99$ and standard deviation $0.03 - 0.04$, irrespective of $\rho$ (Taken from the supporting information of \citet{Guerra12_pnas}).}
\label{Fig4-2}
\end{figure*}

\begin{figure}
\centering
\includegraphics[width=\columnwidth]{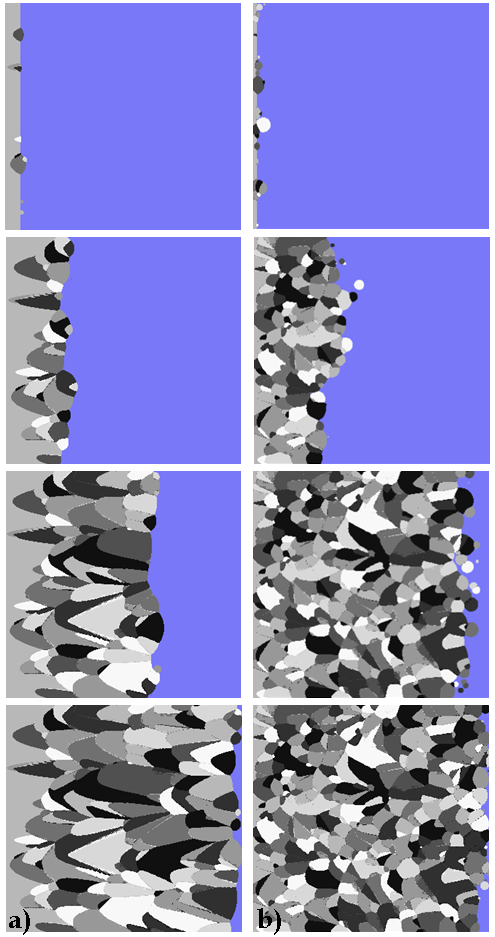}
\caption{Reconstructed sequence of crack propagation and microfailure events at the microscale for two different values of $K$, namely $K=2.53~$MPa$.\sqrt{\mathrm{m}}$ (left) and $K=4.18~$MPa$.\sqrt{\mathrm{m}}$ (right). The blue part corresponds to uncracked material and we ascribed an arbitrary gray level to each of the growing microcrack. From this, one can deduce the intersection points between microcracks that coincide with the observed conics marks on the fracture surfaces. The size of the two zones of analysis is $2.5\times 2.5\un{mm}^2$. The macroscopic crack propagates from left to right.}
\label{Fig4-3}
\end{figure}

\begin{figure}
\centering
\includegraphics[width=0.9\columnwidth]{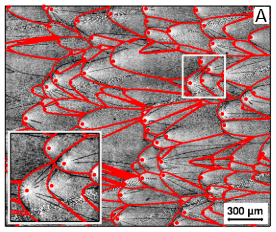}
\includegraphics[width=0.9\columnwidth]{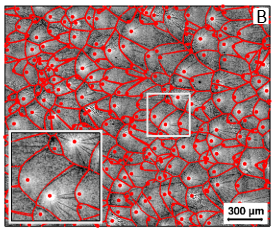}
\caption{Qualification of the procedure: Superimposition of the conic pattern as obtained from reconstruction onto that observed by fractography. A: $K=2.77\un{MPa}.\sqrt{\mathrm{m}}$. B: $K=4.18\un{MPa}.\sqrt{\mathrm{m}}$. Red dots indicate nucleation centers (Taken from the supporting information of \citet{Guerra12_pnas}).}
\label{Fig4-4}
\end{figure}

\begin{figure}
\centering
\includegraphics[width=0.9\columnwidth]{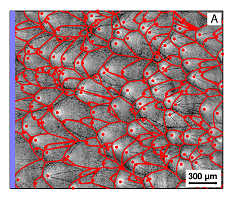}
\includegraphics[width=0.9\columnwidth]{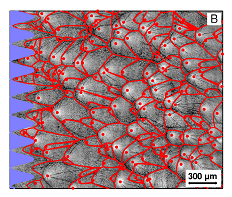}
\includegraphics[width=0.9\columnwidth]{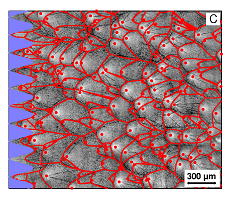}
\caption{Influence of initial crack front morphology on the reconstruction for $\rho =45.0\un{mm}^{-2}$ ($K =3.65~$MPa.$\sqrt{m}$). Three different initial conditions were used. (A) Straight vertical line. (B) Vertical sinusoidal shape with a period of $186~\mu\mathrm{m}$ and a peak-to-peak amplitude of $242~\mu\mathrm{m}$. (C) Same sinusoidal shape, but translated vertically over half a period (Taken from the supplementary information of \citet{Guerra12_pnas}).}
\label{Fig4-5}
\end{figure}
 
We used Eq. \ref{Eq2} to directly extract the ratio $c=c_2/c_1$ for each couple of interacting microcracks ($\sim 400$ couples per image) on different optical pictures obtained at different speeds (Fig. \ref{Fig4-2}). Regardless of the macroscopic crack velocity, we obtain for $c$ a Gaussian distribution centered on 1 with a standard deviation of about 0.03 (Fig. \ref{Fig4-2}). This analysis permits to demonstrate that all microcracks grow at the same velocity $c_m$ inside the DZ. Note that, up to now, nothing prevents $c_m$ to vary with $v$ or $K$. The mechanism that fixes the value for $c_m$ will be discussed later in this paper (see section \ref{sec:6}). By using $c_2 = c_1 = c_m$ in Eq. \ref{Eq2}, the following relation can be obtained:

\begin{align}
\frac{y}{\Delta}= \pm & \left[ 4\frac{d_n(2\Delta-d_n)}{(\Delta-2d_n)^2}\left(\frac{x}{\Delta}\right)^2
-4\frac{d_n(2\Delta-d_n)}{(\Delta-2d_n)^2}\left(\frac{x}{\Delta}\right) \right. \nonumber \\
                      &  \left. \quad +\frac{d_n(2\Delta-d_n)}{(\Delta-2d_n)^2} -4\frac{d_n}{\Delta} +4\left(\frac{d_n}{\Delta}\right)^2 \right]^{1/2},
\label{Eq3}
\end{align}
\noindent where $d_n=(\Delta -c_m\tau)$ represents the critical distance between the front of the incident microcrack and the center of the nucleated microcrack at the instant of its nucleation. This distance is twice the distance between the apex and the focus of the considered microcrack and, as such, $d_n$ can be directly extracted from fractographic images (Fig. \ref{Fig4-1}:C).

We are now in a position to reconstruct the full dynamics of microcracking events from the knowledge of two series of parameters that can be directly extracted from fractography: The position of nucleation sites and the critical nucleation distance $d_n$ associated to each conic mark. We analyzed several areas of the fracture surfaces, corresponding to various macroscopic velocities above the microcracking onset ($v > v_a$). In all cases, nine optical images with partial overlap were recorded and gathered into a single large image -- This ensures adequate statistics in the following analyses. The methodology to extract the data is the following: 
\begin{itemize}
\item We record the coordinates $x$ and $y$ of the nucleation sites; 
\item We estimate the nucleation distance, $d_n$, by inferring, for each nucleation site, who was the parent microcrack from the relative position of the conics apex with respect to the nucleation site (family criterion). Then, $d_n$ is twice the apex-nucleation center distance.
\end{itemize}

Following \citet{Ravichandar97_jmps}, the reconstruction is initiated with an originally straight crack front positioned at the left of the zone of analysis. This front propagates from left to right, at a constant velocity of $1\un{pixel/time step}$. Once the distance between one of the microcrack nucleation sites and the primary crack front reaches $d_n$, a secondary microcrack starts to grow radially at the same velocity. The front is now the combination of both the translating straight front and the radially growing circular front. Beyond a given time, these two fronts intersect and define the position of the conic mark. When the growing front reaches again a distance $d_n$ associated with another site, a new microcrack is nucleated. The procedure is repeated until all the nucleation sites have been triggered.

Figure \ref{Fig4-3} shows two examples of reconstruction for two different macroscopic velocities and Fig. \ref{Fig4-4} superimposes the conic marks obtained by the simulation to those really observed in fractography. We note the fairly good agreement between experiment and reconstruction away from the edges (in the center of the images). This permits to validate the method. Thanks to it, we are able to reconstruct deterministically the dynamics of micro-damage during fast fracture in PMMA from the observation and analysis of the patterns observed on the post-mortem fracture surfaces. We emphasize the resulting resolution of $\sim 1~\mu\mathrm{m}$ (1 pixel) in space, and of $10\un{ns}$ in time (pixel size/$c_m$ with $c_m\approx 200\un{m/s}$ as it will be demonstrated further in the paper). This is far better than what can be obtained via conventional experimental mechanics methods as e.g. acoustic emission or fast imaging methods.

Note however some discrepancies between experiments and reconstructions on the left, top and bottom edges of the analyzed zones. Such finite size problems are unavoidable. Along the top and bottom sides of the image, the influence of microcracks outside the field of view are naturally ignored. Also, a non-realistic straight vertical front has been used, as we cannot predict the precise instants at which the leftmost centers have turned to microcracks during the reconstruction. In order to test the sensitivity of the reconstruction to the initial front shape, we have performed different simulations with the same inputs except the initial front shape. In Fig. \ref{Fig4-5}, three different cases are tested : (i) a straight vertical front, (ii) a sinusoidal front with an amplitude and a period equivalent to the mean value of a wavy front far from the first step in previous reconstruction and (iii) the same sinusoidal front vertically shifted by half a period. We can observe that, except of course at the very beginning of the simulation (i.e. on left side of the images), the difference between these three cases are very small. Moreover, these small differences tend to disappear as the front to propagate, i.e. as more and more microcracks are involved. In some cases, we have observed differences propagating over the whole image. Perfect reconstructions would be reached only by analyzing the whole fracture surfaces, and not some partial areas. Knowing all the history of crack propagation is the only solution for the determination of the real initial front shape which could allow a perfect reconstruction.

\section{Quantitative selection of the microdamaging state by crack loading}\label{sec:5}

The reconstruction method, described in the previous section, shows that the dynamics of micro-damage is entirely determined by the position of nucleation sites, the nucleation distances $d_n$, and the microscopic velocity of the fronts $c_m$. To understand what selects these three parameters, they will be explored in more detail.

\subsection{Spatial distribution of nucleation centers}\label{sec:5.1}

\begin{figure}
\centering
\includegraphics[width=0.9\columnwidth]{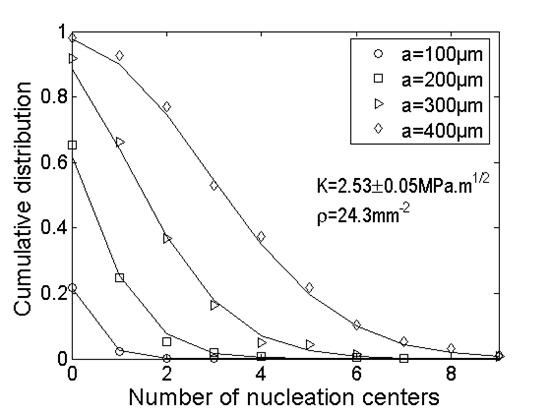}
\includegraphics[width=0.9\columnwidth]{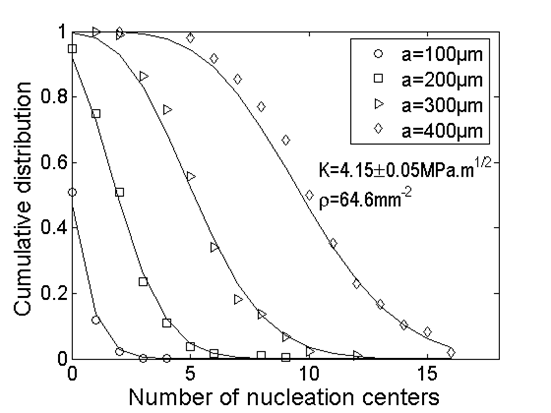}
\caption{Cumulative distribution of the number of nucleation centers contained in square regions of lateral size $a$, for two different fractographic images. Solid lines represent Poisson fits of parameter $\rho a^2$, where the fitted value $\rho$ can be identified with the center density.}
\label{Fig5-1}
\end{figure}

\begin{figure}
\centering
\includegraphics[width=0.9\columnwidth]{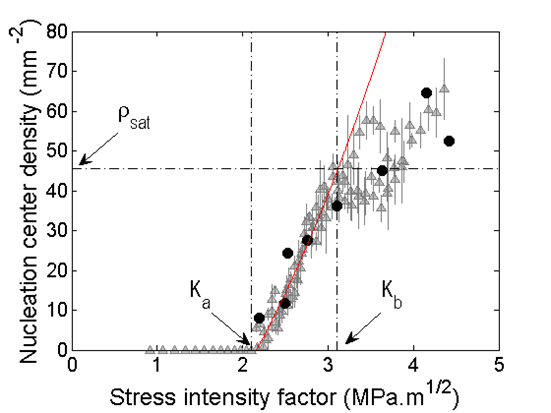}
\caption{Variation of the density $\rho$ of nucleation centers with applied SIF. Black disks correspond to the eight images on which post-mortem reconstruction was performed. Red line is a fit via Eq. \ref{Eq4}. $K_a$ and $K_b$ are associated with microcracking and microbranching onset, respectively (Taken from \citet{Guerra12_pnas}).}
\label{Fig5-2}
\end{figure}

To characterize the spatial distribution of nucleation sites, we computed the distribution of the number of sites that can be observed on a square of a given size $a$. Figure \ref{Fig5-1} shows these distributions for two different fractographic pictures, i.e. for two values of $K$. These distributions follow a Poisson distribution with a single parameter $\rho a^2$ where $\rho$ is constant over a given fractographic picture but varies with $K$. This reflects a homogeneous and uncorrelated distribution of the  sites all along the post-mortem fracture surface. Then, $\rho$ can be identified with the surface density of nucleation sites and fully characterizes their spatial distribution.

Figure \ref{Fig5-2} presents the evolution of $\rho$ with loading $K$. The density is naturally zero for $K$ smaller than $K_a$ (SIF value at microcracking onset, \ie when $v=v_a$) as there is no microcracking. Beyond this value, it increases with $K$. Also, this curve exhibits large fluctuations and seems to saturate when the density reaches a value $\rho_{sat}\approx 45.5\un{mm}^2$, beyond $K_b$. Note that this latter value coincide with microbranching onset and, therefore the $\rho(K)$ curve does not really make sense above $K_b$ since the front has splitted into various microbranches, which prevents us from correlating the macroscopic $K$ value to the number of conic marks on the main crack only.

To account for the curve $\rho(K)$, let us consider \citep{Scheibert10_prl} that a discrete population of weak localized zones is present within the material with a bulk density $\rho_v$. Let us then assume that a microcrack can nucleate in these areas provided the two following  conditions are fulfilled:
\begin{itemize}
\item The local stress at the considered zone is large enough, \ie larger than a given threshold value $\sigma^*$ (smaller than the yield limit $\sigma_Y$ of the material);
\item The considered zone is far enough from the main crack front to allow the microcrack to grow, i.e. at a distance greater than $d_a$.
\end{itemize}
\noindent The density $\rho$ of nucleated microcracks per unit area of fracture is then given by $\rho_v \{h_\perp - d_a\}$ where $h_\perp$ is the thickness of the layer where the stress has exceeded $\sigma^*$ ($h_\perp$ is measured perpendicular to the mean plane of fracture). The singular form of the stress field at crack tip leads to write $h_\perp \propto K^2/{\sigma^*}^2$, which yields \citep{Scheibert10_prl}:
\begin{eqnarray}  
&\rho=0 \quad &\mathrm{for} \quad K \leq K_a \nonumber\\
&\rho \propto K^2-K_a^2 \quad &\mathrm{for} \quad K \geq K_a,
\label{Eq4}
\end{eqnarray}
\noindent with $K_a \propto \sigma^* \sqrt{d_a}$. This equation, plotted in red on Fig. \ref{Fig5-2}, reproduced quite well the experimental data below $K_b$. 

 As the determination of $\rho$ from fracture surface is much more accurate and consensual than the determination of $K$ (based on FEM calculation with inherent approximations), in the following, we will use $\rho$ rather than $K$ as the control parameter as a function of which the various quantities further defined in this paper will be plotted. Indeed, $\rho$ fully characterizes the spatial distribution of nucleation sites and monotonically evolves with $K$ within the relevant range $K_a\leq K \leq K_b$.

\subsection{Chronology of microfailure events}\label{sec:5.2}

\begin{figure}
\centering
\includegraphics[width=0.9\columnwidth]{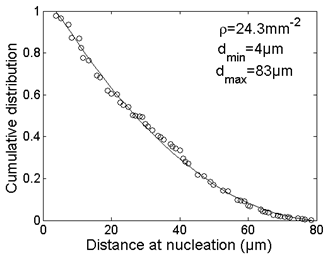}
\includegraphics[width=0.9\columnwidth]{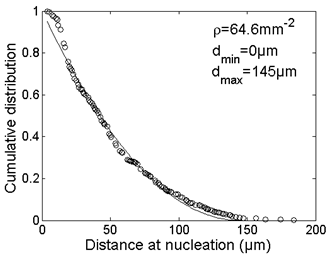}
\caption{Cumulative distribution of the nucleation distances $d_n$, as measured for two different fractographic images. Solid lines are fits with $P(d_n)=\left((d_{max}-d_n)/(d_{max}-d_{min})\right)^2$.}
\label{Fig5-3}
\end{figure}

\begin{figure}
\centering
\includegraphics[width=0.9\columnwidth]{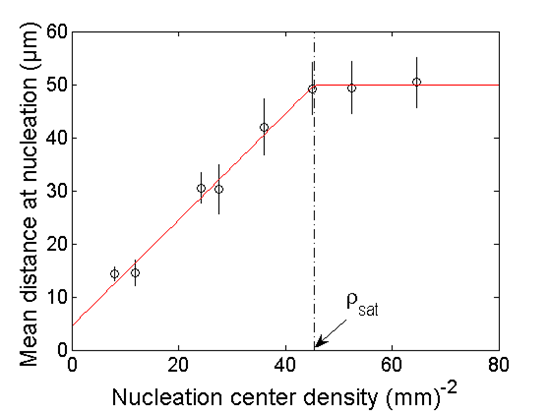}
\includegraphics[width=0.9\columnwidth]{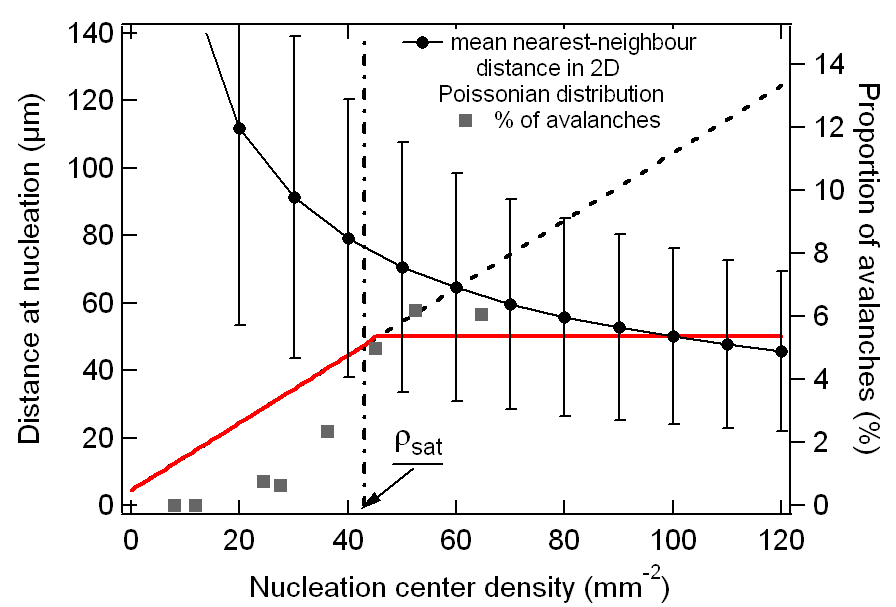}
\caption{Top: Mean distance at nucleation $\bar{d_n}$ as a function of $\rho$. the red curve is a linear fit over $\rho \leq \rho_{sat}=45.5\un{mm}^{-2}$ and a plateau at $\bar{d_n}^{sat}=50~\mu$m over $\rho_{sat}$. Bottom: Saturation of $\bar{d_n}$ and avalanches. The red line is the same as in the top figure. It is compared to the mean nearest-neighbor distance in a Poissonian distribution (black solid line). Solid squares indicate the proportion of microcracks involved in avalanches, as computed from the reconstruction.}
\label{Fig5-4}
\end{figure}

Figure \ref{Fig5-3} shows the cumulative distribution of nucleation distances $d_n$ measured on two typical fracture surfaces. The probability density function (derived from the cumulative distribution) is zero for $d_n\leq d_{min}$, maximum at $d_{min}$, subsequently linearly decreases, and is finally zero as $d_n$ exceeds $d_{max}$. This distribution form was observed for all the fracture surfaces, irrespectively of $\rho$. $d_{max}$ is found to increase with $\rho$, while $d_{min}$ decreases with $\rho$ and remains very close to zero \citep{Guerra12_pnas}. 

The variation of the mean value $\bar{d_n}$ of $d_n$ with $\rho$  exhibits two regimes (Fig. \ref{Fig5-4}:Top): (i) A linear increase with $\rho$ followed by (ii) a saturation ($\bar{d_n}^{sat} \approx 50~\mu\mathrm{m}$) when $\rho$ is greater than the value $\rho_{sat}$ associated in Fig. \ref{Fig5-2} with the microbranching onset (at $K = K_b$). 

To understand the saturation origin, we plotted, on Fig. \ref{Fig5-4}:Bottom,  the evolution of the mean nearest-neighbor distance $\langle \Delta r \rangle$ with $\rho$ as it is predicted for a 2D Poissonian spatial distribution ($\langle \Delta r \rangle=1/2\sqrt{\rho}$). The plotted errorbars also indicate the associated standard deviation ($\sigma_{\Delta r}=\sqrt{(4-\pi)/(4\pi\rho)}$). When $\rho$ reaches $\rho_{sat}$, in few cases, the distance between two neighbouring nucleation centers is of the order of the nucleation distance. As a result, two centers can nucleate almost at the same time (avalanche effect). The number of microcracks involved in such avalanches, also plotted in the figure, increases in the vicinity of $\rho_{sat}$. This strongly suggests that the observed saturation in the $d_n$ evolution (Fig. \ref{Fig5-4}:Top) results from this steric effect. 

\subsection{Velocity of microcrack growth}\label{sec:5.3}

\begin{figure}
\centering
\includegraphics[width=0.85\columnwidth]{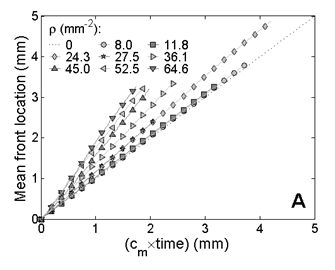}
\includegraphics[width=0.85\columnwidth]{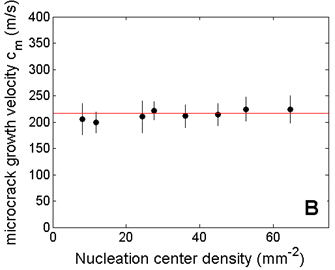}
\caption{Determination of the local propagation velocity $c_m$ of individual microcracks. Left: Evolution of the mean crack front position as a function of $c_m \times t$. Different curves correspond to different fractographic images with different microcrack density. Slope of these curves defines the ratio $A$ between continuum-level scale failure velocity $v$ and microcrack velocity. Right: Variation of $c_m$ with $\rho$. Horizontal red line indicates the mean value $c_m \approx 217$m/s.}
\label{Fig5-5}
\end{figure}

The fractographic analysis performed in Sec. \ref{sec:4} showed that, within the DZ, all microcracks grow at the same velocity $c_m$. Unfortunately, it does not allow us to directly measure the value of $c_m$ and its possible dependence with $\rho$. Indeed, Eq. \ref{Eq2} shows that the form taken by the conical brands depends only on the position of the nucleation sites and on the nucleation distance but not on the value of $c_m$.

We propose to use the deterministic reconstructions obtained in Sec. \ref{sec:4} to connect the macroscopic crack speed $v$ to $c_m$. Figure \ref{Fig5-5}:A shows the evolution of the mean front position measured from reconstructions of different fracture surfaces (i.e. for different densities of nucleation centers) as a function of numerical time step normalized by $c_m$ (homogeneous to a distance, corresponding to the distance travelled by a crack front at a velocity $c_m$ in absence of micracking events). All the curves are linear with a slope $A$ that defines the ratio $v/c_m$. At this point, one should recall that the reconstruction procedure starts with an unrealistic initial straight crack front. We checked that this does not affect the measured value $A$ \citep{Guerra12_pnas}. Then, since $v$ is known, $c_m$ can be deduced and plotted, on Fig. \ref{Fig5-5}:B as a function of $\rho$. This microscopic velocity is found to be a constant, $c_m\approx 217\un{m/s} = 0.24 C_R$, independent of the loading $K$, the density $\rho$, and the macroscale velocity $v$.  The implications of this result will be further discussed in the next section.

\begin{figure}
\centering
\includegraphics[width=0.95\columnwidth]{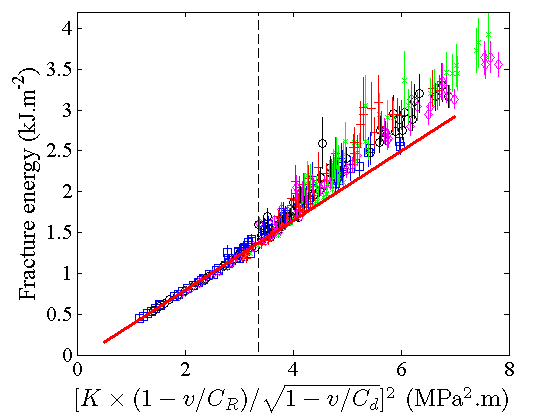}
\caption{Variation of the fracture energy $\Gamma$ with $K_d=K\times (1-v/C_R)/\sqrt{1-v/C_D}$. The vertical dashed line corresponds to microcracking onset $K_d^a=1.83\un{MPa}\sqrt{\mathrm{m}}$. Note that $K_d$ can {\em stricto-senso} be identified with the dynamic SIF in absence of microcracking, \ie for $K_d \leq K_d^a$. The solid red line is a fit with Eq. \ref{Eq5}.} 
\label{Fig5-6}
\end{figure}

The fracture energy measurements $\Gamma(v)$ made in Sec. \ref{sec:3} will now allow us to understand why such a constant velocity $c_m$ is selected within the microcracking regime. Figure \ref{Fig5-6} shows the evolution of $\Gamma$ with the {\em dynamic} SIF $K_d$. In the low speed regime (i.e. for low $K_d$) where crack propagates without involving any microdamaging or microbranching, $\Gamma$ scales with $K_d^2$. We can also expect that the FPZ size $R_c$ scales with $K_d^2$: $R_c = K_d^2/\alpha\sigma_Y^2$ where $\alpha$ is a constant of order $1$, and $\sigma_Y$ is the material's yield stress. The scaling of $\Gamma$ with $K_d$ thus indicates a linear variation between $\Gamma$ and $R_c$ and, following \citet{Scheibert10_prl}, a constant dissipated energy $\epsilon$ per volume unit within the FPZ. The fracture energy then writes:

\begin{equation}
\Gamma=\frac{\epsilon}{\alpha\sigma_Y^2}K_d^2+\left( \frac{1}{E} - \frac{\epsilon}{\alpha\sigma_Y^2}\right)K_c^2
\label{Eq5}
\end{equation}

\noindent As long as $v \leq v_a$ (thus without microdamage), one can relate $K_d$ to $K$ \citep{Freund90_book}: $K_d=K \times (1-v/C_R)/\sqrt{1-v/C_D}$ where $C_D$ refers to the speed of dilational waves (cf. Tab. 1). By replacing $K$ with $K_d$ in Eq. \ref{Eq1}, and by subsequently injecting the result in  Eq. \ref{Eq5}, one gets the following expression for $\Gamma$ vs. $v$:

\begin{equation}
\Gamma=\frac{1-\frac{\epsilon E}{\alpha\sigma_Y^2}}{1-\frac{\epsilon E}{\alpha\sigma_Y^2}\frac{1-v/C_R}{1-v/C_D}}\frac{K_c^2}{E}
\label{Eq6}
\end{equation}

\noindent This expression yields a divergence of $\Gamma$ at a finite velocity $c_{\infty}$ given by:

\begin{equation}
c_\infty= \left(\frac{\epsilon E}{\alpha\sigma_Y^2}-1\right) \frac{C_R C_D}{\frac{\epsilon E}{\alpha\sigma_Y^2}C_D-C_R}
\label{Eq7}
\end{equation}

\noindent In this expression, the only unknown quantity is $\epsilon/\alpha\sigma_Y^2$. This can be evaluated by fitting the first linear part (i.e for $K_d\leq K_d^a$) of the curve $\Gamma$ vs. $K_d^2/E$ plotted in    Fig. \ref{Fig5-6} with Eq. \ref{Eq5}. One then gets: 

\begin{equation}
c_\infty \simeq 204\un{m/s}=0.23C_R
\label{Eq8}
\end{equation}

This value sets the maximum crack growth velocity in the absence of microcracking: Beyond $c_{\infty}$ dissipation diverges within the FPZ. Assuming that the same dissipation mechanisms are involved within the FPZ of each microcrack when $v > v_a$, we expect that the limiting speed of microcracks is also determined by $c_{\infty}$, so that $c_m \approx c_{\infty}$.

\section{Role of microdamaging in the selection of continuum-level scale failure velocity}\label{sec:6}

\begin{figure}
\centering
\includegraphics[width=0.9\columnwidth]{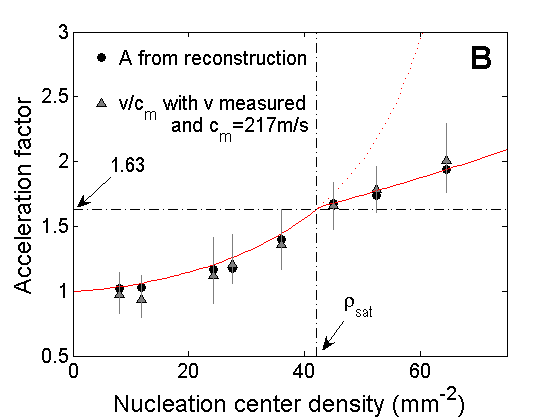}
\caption{Acceleration factor $A=v/c_m$ as a function of the microcrack density $\rho$. Red line is a fit by Eq. \ref{Eq9} with $b=1.19\pm0.02$. Horizontal dot line defines $v_b/c_m$ and passes through the slope breakdown observed when $\rho=\rho_{sat}$ (Taken from \citet{Guerra12_pnas})}
\label{Fig6-1}
\end{figure}

The previous section has allowed us to better understand how the different variables that characterize the microdamage developing within the DZ are selected when $v$ exceeds $v_a$. The most surprising observation concerns the propagation velocity of  microcracks $c_m$ that remains constant, of the order of $217\un{m/s}$, and significantly lower than the macroscopic velocity $v$. In other terms, damage does not slow down the macroscopic crack tip as it was commonly believed until now \citep{Ravichandar04_book,Ravichandar97_jmps,Washabaugh94_ijf}, but on the contrary it accelerates it. And this acceleration factor is all the more important as $\rho$ (or equivalently $K$) increases (Fig. \ref{Fig6-1}).

\begin{figure}
\centering
\includegraphics[width=0.9\columnwidth]{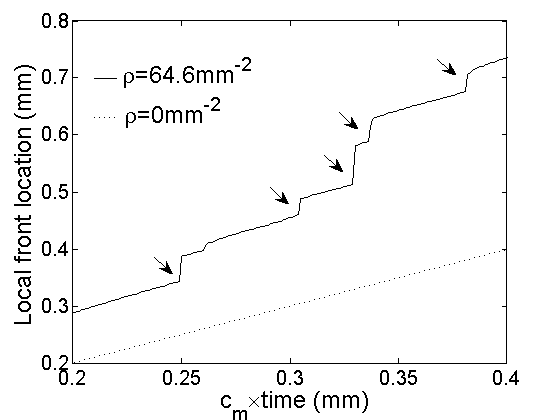}
\caption{Position of a given point of the front as a function of "time" $c_m \times t$ as obtained from the reconstructed simulation, applied to a fracture surface with $\rho=64.6\un{mm}^{-2}$. The dotted straight line indicates the expected position without microdamage (slope 1). The jumps (arrows) correspond to coalescence events with microcracks (Taken from \citet{Guerra12_pnas})}
\label{Fig6-2}
\end{figure}

The temporal evolution of a given point on the main front, shown in Fig. \ref{Fig6-2}, permits to better understand the effect of microdamage on the dynamics of macroscopic cracking. The trajectory of this point exhibits random and sudden jumps. These jumps actually correspond to the coalescence of a microcrack with the main front. Between these events, the local front velocity remains close to $c_m$ (\ie the slope is equal to 1). Such an acceleration effect can be captured by a simple mean-field model, detailed in \citet{Guerra12_pnas}, which simply consists in counting the number of coalescence events as the main front propagates over a unit of length when the nucleation sites are located at the nodes of a square lattice. One then gets:

\begin{equation}
A = \frac{1}{1-b \bar{d_n} \sqrt{\rho}}
\label{Eq9}
\end{equation}

where $b$ is a numerical factor close to 1. This equation, combined with the evolution of $\bar{d_n}$ with $\rho$ (red line in Fig. \ref{Fig5-4}:Top) is plotted in red in Fig. \ref{Fig6-1}. It reproduces quite well both the data coming from experiments (grey triangles) and from the reconstruction (black circle). Note the change in slope observed at $\rho_{sat}$. At this point, $A=A_b=1.63$. This corresponds to a ratio $A_b \simeq v_b/c_m$ where $v_b\simeq 0.4C_R$ is associated with the microbranching onset. This strongly suggests that the steric effect at the origin of the saturation of $d_n$ also plays a role (still unsolved) in the micro-branching instability.

\section{Conclusion}\label{sec:7}

The series of experiments reported here were designed to shed light on the dissipation mechanisms that develop during fast crack growth in PMMA. They reveal that, above a well-defined critical velocity, PMMA stops to behave {\em stricto senso} as a nominally brittle material and crack propagation goes hand in hand with microcracks that nucleate and grow ahead of the main front. Those let characteristic marks (referred to as conic marks) on post-mortem fracture surfaces, the morphological analysis of which allowed us to reconstruct the full dynamics of microfailure events. The simultaneous space and time accuracies were of the order of the micrometer and the tens on nanoseconds, \ie far better than what was reachable up to now.   

Analysis of the reconstructions demonstrated that the true local propagation velocity of single cracks remains limited to a fairly low value $c_m \simeq 0.23 c_R$ in PMMA, while the {\em apparent} fracture velocity measured at the continuum-level scale (\eg via potential drop method) can be much higher (up to twice larger!). Such an anomalously high measured velocity results in fact from the coalescence of microcracks with each other and with the main front -- all of them growing at $c_m$. In other words, the main effect of microdamaging is not, as commonly believed up to now \citep{Washabaugh94_ijf,Ravichandar04_book, Ravichandar97_jmps},  to slow down fracture by increasing dissipation within the DZ, but on the contrary to boost the propagation of the main crack.

The value $c_m$ that limits the true local velocity of single crack fronts in PMMA in the dynamic regime is set by the physico-chemical dissipative and non-linear processes that develop within the FPZ, \eg thermal \citep{Estevez00_jmps}, viscoelastic effects \citep{Boudet96_jpii, Persson05_pre} or hyperelastic processes \citep{bouchbinder08_prl,Livne10_science, Buehler03_nature}. Hence, it will depend on the considered material. On the other hand, the boost effect induced by microdamaging takes the form of a purely geometrical factor controlled by two quantities: the density $\rho$ of nucleation sites and the mean distance $\overline{d_n}$ at nucleation. These two variables fully characterize the damaging state within DZ and evolve with $K$. Once ascribed, they permit to relate the continuum-level fracture speed $v$ to the true local propagation speed $c_m$. Further work is required to unravel how PMMA internal structure selects the material-dependent quantities $c_m$, $\rho(K)$, and $\overline{d_n}$. 

The presence of microcracks forming ahead of a dynamically growing crack has been evidenced in a variety of materials, \eg in most brittle polymers \citep{Ravichandar98_ijf,Du10_jms}, in rocks \citep{Ahrens93_jgr}, in some nanophase ceramics and nanocomposites \citep{Rountree02_armr}, in oxide glasses \citep{Rountree10_pcge}, in metallic glasses \citep{Murali11_prl}, etc. The boost mechanism demonstrated here on PMMA is expected to hold in this whole class of materials. Note that a similar effect has also been invoked \citep{Prades05_ijss} in oxide glasses under stress corrosion, where ultra-slow crack propagation (down to few tenths of nanometers per second) was  found  to involves "nano"-cracks forming ahead of the main crack tip \citep{Celarie03_prl,Celarie03_ass,Bonamy06_ijf,Ferretti11_jmps}. This yields us to conjecture that failures with anomalously high apparent velocities measured at continuum-level scale may arise in all situations involving propagation-triggered microcracks, including \eg shear fracture in compressed granite \citep{Moore95_jsg} , thermal failure in shale \citep{Kobchenko11_jgr,Panali12_spe}, and more generally failure of so-called quasi-brittle materials. One of the most robust observations in this field is the power-law form followed by the distribution in size and in time interval between two successive microfailure events (see \citep{Bonamy09_jpd,Deschanel09_jpd} for recent reviews). It would be interesting to see whether or not the microcracking evidenced here in dynamic failure regime share the same scale-free features and how this affects the boost effect. Work in this direction is under way.

\begin{acknowledgements}
We warmly thank K. Ravi-Chandar (Univ. of Texas, Austin) for many illuminating discussions. We also thank T. Bernard (SPCSI) for technical support, P. Viel and M. Laurent (SPCSI) for gold deposits, and A. Prevost (ENS, Paris) for his help with the profilometry measurements at ENS. We also acknowledge funding from French ANR through Grant No. ANR-05-JCJC-0088 and from Triangle de la Physique through Grant No. 2007-46.
\end{acknowledgements}


\end{document}